\documentstyle[sprocl]{article}
\def\be{\begin{equation}}
\def\ee{\end{equation}}
\def\bea{\begin{eqnarray}}
\def\eea{\end{eqnarray}}
\begin{document}
\title{SOFT SUPERSYMMETRY--BREAKING TERMS FROM SUPERGRAVITY AND 
SUPERSTRING MODELS\footnote{To appear in the book `Perspectives on 
Supersymmetry', World Scientific, Editor G. Kane.}}
\author{ A. BRIGNOLE }
\address{Theory Division, CERN, CH-1211 Geneva 23, Switzerland}
\author{ L.E. IB\'A\~NEZ  
}
\address{Departamento de F\'{\i}sica Te\'orica, 
Universidad Aut\'onoma de Madrid\\
Cantoblanco, 28049 Madrid, Spain}
\author{ C. MU\~NOZ\footnote{On leave of absence from Departamento de
F\'{\i}sica Te\'orica, Universidad Aut\'onoma de Madrid, 
Cantoblanco, 28049 Madrid,
Spain.} }
\address{ Department of Physics,
Korea Advanced Institute of Science and Technology\\ 
Taejon 305-701,
South Korea}
\maketitle\abstracts{
We review the origin of soft supersymmetry--breaking 
terms in $N=1$ supergravity 
models of  particle physics.  We first consider general formulae for those
terms in general models with a hidden sector
breaking supersymmetry at an intermediate energy scale.  The results
for some simple models are given. We then consider the results 
obtained  in some simple superstring  models in which particular
assumptions about the origin of supersymmetry breaking are made. These 
are models in which the seed of supersymmetry
breaking is assumed to be originated in
the dilaton/moduli sector of the theory.
}

\thispagestyle{empty}

%\vskip-17.cm
\rightline{}
\rightline{CERN--TH/97--143}
\rightline{FTUAM 97/7}
\rightline{KAIST--TH--97/7}
\rightline{hep-ph/9707209}
\vskip2in

\newpage
\setcounter{page}{1}

\section{Introduction }
%\vspace{-0.7cm}
%\subsection{Typeset Scripts}
%\vspace{-0.35cm}

A phenomenological implementation of the idea of supersymmetry (SUSY) 
in the
standard model requires the presence of SUSY breaking.  
There are essentially two large families  of models in this context, 
depending on whether the scale of spontaneous SUSY breaking is 
high (of order  $10^{10}$--$10^{13}$~GeV) or low (of order 1--$10^2$~TeV).
We will focus on the former possibility. 
The latter possibility, considered in other chapters of
this book,  has only recently received sufficient attention,
since it was realized from the very first days of SUSY phenomenology
that the  existence of  certain supertrace constraints in
spontaneously  broken SUSY theories  made the building of 
realistic models quite complicated. Possible solutions to these
early difficulties are discussed elsewhere and we are not going to 
discuss them further here.

A more pragmatic attitude to the issue of SUSY breaking is the 
addition of explicit soft SUSY-breaking terms of the appropriate size
(of order $10^2$--$10^3$~GeV) in the Lagrangian and 
with appropriate flavour 
symmetries to avoid dangerous flavour-changing neutral currents (FCNC) 
transitions. The problem with this 
pragmatic attitude is that, taken by itself, lacks any theoretical 
explanation. Supergravity theories provide an attractive context
that can justify such a procedure. Indeed, if one considers the 
SUSY standard model and couples it to $N=1$ supergravity, 
the spontaneous breaking of local SUSY in a hidden sector
generates explicit soft SUSY-breaking terms of the required form 
in the effective low-energy Lagrangian \cite{HPN,CM}.  
If SUSY is broken at a scale  $\Lambda _S$, the 
soft terms have a scale of order $\Lambda _S^2/M_{Planck}$. Thus 
one obtains the required size if SUSY is broken at an 
intermediate scale $\Lambda _S \sim  10^{10} $~GeV, as mentioned above. 
Large classes of supergravity models, as we discuss in section 2,
give rise to universal soft SUSY-breaking terms, providing for an 
understanding of FCNC supression. 
In the last few years it has often 
been  stated in the literature that this class of supergravity  models 
have a flavour-changing problem.  We think more appropriate
to  say that  
some particular models get interesting {\it constraints}  from
FCNC bounds. A generic statement like that seems unjustified,
since it is usually based on a strong assumption, i.e. the existence 
of a region in between the grand unified theory (GUT) 
scale and the Planck (or superstring) scale 
in which important flavour non-diagonal renormalization effects take place. 

Recently there have been studies of supergravity models
obtained in particularly simple classes of 
superstring compactifications \cite{CM2}.
Such heterotic models have a natural hidden sector built-in: the
complex dilaton field $S$ and the complex moduli fields $T_i$. 
These gauge singlet fields are generically present in four-dimensional
models: the dilaton arises from the gravitational sector of the theory
and the moduli parametrize the size and shape of the compactified
variety.
Assuming that 
the auxiliary fields of those multiplets are the seed of SUSY breaking,
interesting  predictions for this simple class of models are obtained. 
These 
are reviewed in section 3. The analysis does not assume
any specific SUSY-breaking mechanism. 
We leave section 4 for some final comments 
and additional references to recent work.

\section{Soft terms from supergravity }

\subsection{General computation of soft terms}

\noindent The full N=1 supergravity Lagrangian \cite{HPN} 
(up to two derivatives) is specified in terms of two functions
which depend on the chiral superfields $\phi_M$ of the theory
(denoted by the same symbol as their scalar components): the
analytic gauge kinetic function $f_a(\phi_M)$ and the real
gauge-invariant K\"ahler function $G(\phi_M,\phi^*_M)$.
$f_a$ determines the kinetic terms for the fields in the vector multiplets
and in particular the gauge coupling constant, $Re f_a=1/g_a^2$. The
subindex $a$ is associated with the different gauge groups of the 
theory since
in general ${\cal G}=\prod_a {\cal G}_a$. 
For example, in the case of the pure SUSY
standard model 
coupled to supergravity, $a$ would correspond 
to $SU(3)_c$, $SU(2)_L$, $U(1)_Y$.
$G$ is a combination of two functions
\begin{eqnarray}
G(\phi_M,\phi^*_M)=K(\phi_M,\phi^*_M)+\log|W(\phi_M)|^2 \ ,
\label{G}
\end{eqnarray}
where $K$ is the K\"ahler potential, $W$ is the complete analytic
superpotential, and we use from now on the standard supergravity 
mass units
where $M_P \equiv M_{Planck}/\sqrt{8\pi}=1$.
$W$ is related with the Yukawa couplings (which eventually determine the
fermion masses) and also includes possibly 
non-perturbative effects
\begin{eqnarray}
W={\hat W(h_m)} +\frac{1}{2}\mu_{{\alpha}{\beta}}(h_m){C}^{\alpha}C^{\beta} 
+ \frac{1}{6} Y_{{\alpha}{\beta}{\gamma}}(h_m){C}^{\alpha}C^{\beta}C^{\gamma} 
+... \ ,
\label{F4}
\end{eqnarray}
where we assume two different types of scalar fields 
$\phi_M=h_m, C^{\alpha}$: 
$C^{\alpha}$ correspond to the observable sector and in particular
include the SUSY standard model fields, 
while $h_m$ correspond to a hidden sector.
The latter fields may develop large ($\gg M_W$) vacuum expectation
values (VEVs) and are responsible
for SUSY breaking if some auxiliary components $F^m$ (see below) 
develop nonvanishing 
VEVs. The ellipsis indicates terms of higher order in $C^{\alpha}$
whose coefficients are suppressed by negative powers
of $M_{P}$.
The second derivative of $K$ determines
the kinetic terms for the fields in the chiral supermultiplets and is
thus important for obtaining the proper normalization of the fields.
Expanding in powers of
$C^{\alpha}$ and 
${C^*}^{\overline{\alpha}}$ we have 
\begin{eqnarray}
K &=&
{\hat K}(h_m,h_m^*)
+\ {\tilde K}_{{\overline{\alpha}}{\beta}}(h_m,h_m^*)
{C^*}^{\overline{\alpha}} C^{\beta }\ 
\nonumber\\ &&
+\ \left[ \frac{1}{2} Z_{{\alpha }{ \beta }}(h_m,h_m^*){C}^{\alpha}
C^{\beta }\ +\ h.c. \ \right]+...\ , 
\label{F3}
\end{eqnarray}
where the ellipsis indicates terms of higher order in
$C^\alpha$ and ${C^*}^{\overline{\alpha}}$.
Notice that the coefficients ${\tilde K}_{{\overline{\alpha}}{\beta}}$,  
$Y_{\alpha \beta \gamma}$, $\mu_{\alpha \beta}$, 
and $Z_{\alpha \beta}$ which appear in (\ref{F4}) and (\ref{F3}) 
may depend on the hidden sector fields in general.
The bilinear terms associated with $\mu_{\alpha \beta}$ 
and $Z_{\alpha \beta}$ are often forbidden by gauge invariance
in specific models, but they may be {\it relevant} in order to solve the
so-called $\mu$ problem in the context of the minimal supersymmetric
standard model (MSSM), as we will discuss below.
In this case the two Higgs doublets, which are necessary to break the 
electroweak symmetry, have opposite hypercharges. Therefore those 
terms are allowed
%\cite{GM,Kaplunovsky} 
and may generate both the $\mu$ parameter
and the corresponding soft bilinear term.

The (F part of the) tree-level supergravity 
scalar potential, which is crucial 
to analyze the breaking of SUSY, is given by
\begin{eqnarray}
& V(\phi _M, \phi ^*_M)\ =\
e^{G} \left( G_M{K}^{M{\bar N}} G_{\bar N}\ -\ 3\right) =  
\left( {\bar F}^{\bar N}{K}_{{\bar N}M} F^M\ -\ 3e^{G}\right)
\ , &
\label{F1}
\end{eqnarray}
%-\ {\overline{F}}^{\overline{m}} \left( \partial_{\overline{m}}\partial_n
%
where $G_M \equiv \partial_M G \equiv \partial G/ \partial \phi_M$ 
and the matrix $K^{M{\bar N}}$ is the inverse of the K\"ahler metric
$K_{{\bar N }M}\equiv{\partial}_{\bar N}{\partial }_M K$.
We have also written $V$ as a function of the $\phi_M$  auxiliary fields, 
$F^M=e^{G/2} {K}^{M {\bar P}} G_{\bar P}$. When, at the minimum of the
scalar potential, some of the hidden sector
fields $h_m$ acquire VEVs in such a way that at least one of their 
auxiliary fields
(${\hat K}^{m {\overline{n}}}$ is the inverse of 
the hidden field metric
${\hat K}_{{\overline {n}} m}$)
\begin{eqnarray}
F^m=e^{G/2} {\hat K}^{m{\overline n}} G_{{\overline n}} 
\label{auxiliary}
\end{eqnarray} 
is non-vanishing, then SUSY is spontaneously broken and soft SUSY-breaking
terms are generated in the observable sector. 
%The matrix
%${\hat K}^{m {\overline{n}}}$ is the inverse of 
%the moduli metric
%${\hat K}_{{\overline {n}} m}$.
Let us remark that,
for simplicity, we are assuming vanishing
D-term contributions to SUSY breaking. When this is not the case,
their effects on soft terms can 
be found e.g. in \cite{KKK}.
The goldstino, which is a combination of the fermionic partners
of the above fields, is swallowed by the gravitino via the superHiggs
effect.
%\cite{cremmer}
The gravitino becomes massive and its mass
\begin{eqnarray}
m_{3/2}=e^{G/2}
\label{gravitino}
\end{eqnarray}
sets the overall scale of the soft parameters.

\vspace{0.1cm}

\noindent {\it General results}

\vspace{0.1cm}

\noindent
Using the above information, the soft SUSY-breaking terms in the 
observable
sector can be computed. 
They are obtained by replacing $h_m$ and their auxiliary fields $F^m$ 
by their VEVs in the supergravity Lagrangian
and taking the so-called flat limit
%\cite{flat} 
where 
$M_{P}\rightarrow \infty$
but $m_{3/2}$ is kept fixed. Then the non-renormalizable 
gravity corrections are formally eliminated and
one is left with a global SUSY
Lagrangian plus a set of soft SUSY-breaking terms. 
On the one hand, from the fermionic part of
the supergravity 
Lagrangian,
%\cite{Sugra} 
soft
gaugino masses for the canonically {\it normalized} gaugino fields
can be obtained 
\begin{eqnarray}
M_a=\frac{1}{2}\left(Ref_a\right)^{-1} F^m \partial_m f_a  \; ,
\label{F2}
\end{eqnarray}
as well as the {\it un-normalized} Yukawa couplings of the observable sector
fermions and 
the SUSY {\it un-normalized} masses of some of them 
(those with bilinear terms either in the superpotential or in the K\"ahler 
potential, e.g. the Higgsinos in the case of the MSSM) 
\begin{eqnarray}
Y'_{\alpha\beta\gamma} &=& 
\frac{ {\hat W}^*}{|{\hat W}|} e^{\hat K/2} {Y}_{\alpha \beta \gamma} \ ,
\label{fermion}
\\
{\mu}'_{\alpha \beta} &=& 
\frac{ {\hat W}^*}{|{\hat W}|} e^{\hat K/2} {\mu}_{\alpha \beta}
+ m_{3/2} Z_{\alpha \beta} - 
{\overline {F}}^{\overline{m}} \partial_{\overline{m}} Z_{\alpha \beta} \ .
\label{fermions}
\end{eqnarray}
On the other hand, scalar soft terms arise from the expansion of
the supergravity scalar potential (\ref{F1})
\begin{eqnarray}
{V}_{soft}=
{m}'^2_{{\overline {\alpha}}{ \beta}}
{C^*}^{\overline{\alpha}} C^{\beta }
+\left( \frac{1}{6} A'_{\alpha \beta \gamma} 
	    {C}^{\alpha}{C}^{\beta}{C}^{\gamma}
  +\frac{1}{2} B'_{\alpha \beta} C^{\alpha}C^{\beta} + h.c.\right) \ .
\label{potencial}
\end{eqnarray}
In the most general case, when hidden and observable sector matter metrics
are not diagonal, the {\it un-normalized} soft scalar masses, trilinear
and bilinear parameters are given respectively by \cite{SW}
\begin{eqnarray}
{m}'^2_{{\overline{\alpha }}{ \beta }} &=& 
\left(m_{3/2}^2+V_0\right){\tilde K_{{\overline{\alpha }}{ \beta }}}
\nonumber\\ &&
-\ {\overline{F}}^{\overline{m}} \left( \partial_{\overline{m}}\partial_n
{\tilde K_{{\overline{\alpha }}{ \beta }}}
-\partial_{\overline{m}} {\tilde K_{{\overline{\alpha }}{ \gamma}}}
{\tilde K^{{ \gamma} {\overline{\delta}} }}
\partial_n {\tilde K_{{\overline{\delta}}{ \beta}}}  \right) F^n \ , 
\label{mmatrix}
\\
A'_{\alpha\beta\gamma} &=& 
\frac{ {\hat W}^*}{|{\hat W}|} e^{{\hat K}/2} F^m \left[ 
{\hat K}_m Y_{\alpha\beta\gamma} + \partial_m Y_{\alpha\beta\gamma}\right. 
\nonumber\\ &&
\left.-\ \left( {\tilde K^{{ \delta} {\overline{\rho}} }}
\partial_m {\tilde K_{{\overline{\rho}}{ \alpha}}} Y_{\delta\beta\gamma}
+\left(\alpha \leftrightarrow \beta \right)+
\left(\alpha \leftrightarrow \gamma\right)\right)
\right]\  ,
\label{mmatrix2}
\\
B'_{\alpha\beta} &=& \frac{ {\hat W}^*}{|{\hat W}|}
e^{{\hat K}/2}\left\{ F^m \left[ 
{\hat K}_m \mu_{\alpha\beta}
+ \partial_m \mu_{\alpha \beta}\right.\right.
\nonumber\\ && 
\left.\left.-\ \left(
{\tilde K^{{ \delta} {\overline{\rho}} }}
\partial_m {\tilde K_{{\overline{\rho}}{ \alpha}}} \mu_{\delta \beta}
+(\alpha \leftrightarrow \beta)\right)\right]
- m_{3/2} \mu_{\alpha\beta}\right\}
\nonumber\\ &&
+\ 
\left(2m_{3/2}^2+V_0\right) {Z}_{\alpha \beta} -    
m_{3/2}{\overline{F}}^{\overline{m}} 
\partial_{\overline{m}} Z_{\alpha \beta} 
\nonumber\\ &&
+\ 
m_{3/2} F^m \left[ \partial_m Z_{\alpha \beta} - \left(
{\tilde K^{{ \delta} {\overline{\rho}} }}
\partial_m {\tilde K_{{\overline{\rho}}{ \alpha}}} Z_{\delta \beta}
+(\alpha \leftrightarrow \beta)\right)\right]
\nonumber\\ &&
-\ {\overline{F}}^{\overline{m}} F^n \left[ \partial_{\overline{m}}
\partial_n 
Z_{\alpha \beta} - \left(
{\tilde K^{{ \delta} {\overline{\rho}} }}
\partial_n {\tilde K_{{\overline{\rho}}{ \alpha}}} \partial_{\overline{m}} 
Z_{\delta \beta}
+(\alpha \leftrightarrow \beta)\right)\right]
\ ,  
\label{mmatrix3}
\end{eqnarray}
where ${\tilde K}^{\alpha {\overline{\beta}}}$ is the inverse of 
the observable matter metric
${\tilde K}_{{\overline {\beta}} \gamma}$.
$V_0$ is the VEV of the scalar potential (\ref{F1}), i.e. the 
tree-level cosmological constant
\begin{eqnarray}
V_0=
{\overline {F}}^{{\overline m}}{\hat K}_{{\overline m}n} F^n - 3m_{3/2}^2
\ .
\label{cc}
\end{eqnarray}
It has a bearing on measurable quantities like scalar masses and therefore
it is preferable to leave it undetermined (see 
\cite{BIM,CKN,FKZ} for a discussion
on this point).
Notice that, 
after normalizing the fields to get
canonical kinetic terms, the first piece in
(\ref{mmatrix}) will lead to universal diagonal soft
masses but the second piece will generically induce
{\it off-diagonal} contributions.
Actually, universality is a desirable property not only to reduce the 
number of independent parameters in SUSY models, but also for 
phenomenological
reasons, particularly to avoid FCNC. 
The latter is a {\it low-energy}  phenomenological constraint that must be 
satisfied by any
supergravity model. It is worth mentioning in this context that one--loop
corrections to the soft parameters 
(\ref{F2}), (\ref{mmatrix}), 
(\ref{mmatrix2}) and (\ref{mmatrix3}) 
%and the $\mu$ term (\ref{fermions}) 
have recently been computed in
\cite{kiwoon}. They may induce FCNC phenomena even when the tree-level 
computation gives a universal soft mass. Also a discussion about the
loop effects on the contribution of the cosmological constant to the
soft terms can be found there.
%Also they may provide several 
%new sources of the $\mu$ term apart from the ones discussed below.
Concerning the 
$A$ and $B$ parameters, notice that we have not
factored out the Yukawa couplings and mass terms respectively as usual, since
proportionality is not guaranteed in (\ref{mmatrix2}) and (\ref{mmatrix3}).
%Indeed, although the first term in
%(\ref{mmatrix2}) (13 ????) is always proportional
%in flavour space to the corresponding Yukawa
%coupling, the same thing is not necessarily true
%for the other terms.
%We will see below that this is precisely what occurs in general with 
%scalar masses and trilinear parameters in
%CY compactifications. However, there is an interesting limit where
%universality is achieved.
Finally, from (\ref{auxiliary}) and (\ref{gravitino}), one can easily
see that the (tree-level) soft parameters in (\ref{F2}), (\ref{mmatrix}) 
and (\ref{mmatrix2}) are generically  ${\cal O}(m_{3/2})$,
as mentioned above. A departure from this result is only possible 
when some of them vanish. We will mention below some examples of 
this type, i.e. the case of no-scale supergravity models and 
special limits of superstring models.

\vspace{0.1cm}

\noindent {\it The $\mu$ problem} 

\vspace{0.1cm}

\noindent 
The set of mass parameters in the MSSM Higgs potential includes, 
besides ${\cal O}(m^2_{3/2})$ soft masses, 
the square of $\mu'_{\alpha \beta}$ 
(\ref{fermions}) and the $B$ parameter (\ref{mmatrix3}), which is 
${\cal O}(m_{3/2}\mu'_{\alpha\beta})$. In order to have correct electroweak
symmetry breaking, the SUSY mass term $\mu'_{\alpha\beta}$ should also be
${\cal O}(m_{3/2})$. This is the so-called $\mu$ problem \cite{CM}. 
In this respect, notice that $\mu'_{\alpha\beta}={\cal O}(m_{3/2})$
is {\it naturally} achieved in the presence of a non-vanishing 
$Z_{\alpha\beta}$ in the K\"ahler potential \cite{GM,JACM}. 
The other possible source
of the mass $\mu'_{\alpha\beta}$ is the SUSY mass $\mu_{\alpha\beta}$
in the superpotential \cite{KN}. This case is more involved since in principle
the natural scale of $\mu_{\alpha\beta}$ would be $M_{P}$.
However, a possible solution can be obtained if the superpotential
contains e.g. a non-renormalizable term \cite{JACM,KN}
\begin{eqnarray}
\lambda(h_m){\hat W}(h_m)H_1H_2\ , 
\label{mu2}
\end{eqnarray}
characterized
by the coupling $\lambda$, which mixes the observable sector with the
hidden sector. Since $m_{3/2}=e^{G/2}=e^{{\hat K}/2}|{\hat W}|$,
if that term exists then an effective $\mu$ parameter ${\cal O}(m_{3/2})$ 
is generated {\it dynamically} when $h_m$ acquire VEVs:
\begin{eqnarray}
\mu=\lambda(h_m){\hat W}(h_m)\ .
\label{mu}
\end{eqnarray}
We should add that both mechanisms to generate $\mu'_{\alpha\beta}$, 
a bilinear term in the K\"ahler potential or in the superpotential, 
could be present simultaneously. 
Notice also that the two mechanisms are equivalent if $Z$ depends only
on $h_m$ (not on $h_m^*$). Indeed, in that case
the supergravity theory is equivalent to the one with a K\"ahler potential
$K$ without the terms $ZH_1H_2+h.c.$ and a superpotential $We^{ZH_1H_2}$,
since the $G$ function (\ref{G}) is the same for both. After expanding
the exponential, the superpotential will have a contribution
$Z{\hat W}H_1H_2$, i.e. a term of the type (\ref{mu2}).
Finally, let us mention that several new sources of the $\mu$ term
due to loop effects on (\ref{fermions}), which are naturally of order
the weak scale, have recently been computed in \cite{kiwoon}. 

We recall that the solutions mentioned here in order to 
solve the $\mu$ problem are {\it naturally} present in superstring models.
For instance, in large classes of superstring models the K\"ahler
potential does contain bilinear terms analytic in the observable fields
as in (\ref{F3}), with specific coefficients $Z_{\alpha\beta}$ 
\cite{KL,LLM,AGNT},
so that a $\mu$ parameter may be naturally generated.
Concerning superpotential contributions, we recall that a `direct' 
$\mu H_1H_2$ term in $W$ (\ref{F4}) is naturally absent 
(otherwise the natural scale for $\mu$ would be $M_{P}$), since in 
supergravity 
models deriving from superstring theory mass terms for light fields are 
forbidden in the superpotential by scale invariance of the theory. 
However, the superpotential (\ref{F4}) may well contain an `effective'
$\mu H_1 H_2$ term, e.g. a term of the type (\ref{mu2}) 
\cite{JACM,AGNT} induced by non-perturbative SUSY-breaking 
mechanisms like gaugino-squark condensation in the hidden sector.

\vspace{0.1cm}

\noindent {\it The low-energy spectrum}

\vspace{0.1cm}

\noindent
The results (\ref{F2}), (\ref{fermion}), (\ref{fermions}),                     
(\ref{mmatrix}), (\ref{mmatrix2}) and (\ref{mmatrix3})         
should be understood as being valid at some high scale ${\cal O}(M_P)$
and the standard RGEs must be used to obtain the low-energy values.
Although the SUSY spectrum will depend in general on the 
details of $SU(2)_L\times U(1)_Y$ breaking, there are several 
particles whose mass is rather independent of those details 
and is mostly given by the boundary conditions and the renormalization 
group running. In particular, 
neglecting all Yukawa couplings except the one of the top, that is
the case of the gluino $g$, all the squarks (except stops and left sbottom)
$Q_L=(u_L,d_L), u_L^c, d_L^c$ and all the sleptons $L_L=(v_L,e_L), e_L^c$.
For all these particles
one can write explicit expressions for the masses in terms of the 
soft parameters (after normalizing the fields to get canonical kinetic
terms). For instance, assuming that gauginos have a common initial
mass (e.g. due to a universal $f$ function) and that there is nothing 
but the MSSM in between the weak scale and the Planck scale, 
one obtains the approximate numerical expressions:  
\begin{eqnarray}
M_g^2(M_Z) \ &\simeq& 9.8 \ M^2 \ ,
%\label{F00004}\\
\nonumber\\
m_{Q_L}^2(M_Z) \ &\simeq& m^2_{Q_L} + 8.3 \ M^2
\ ,
%\label{F00005}\\
\nonumber\\
m_{u^c_L,d^c_L}^2(M_Z) \ &\simeq& m^2_{u^c_L,d^c_L} + 8 \ M^2 \ ,
%\label{F00006}\\
\nonumber\\
m_{L_L}^2(M_Z) \ &\simeq& m^2_{L_L} + 0.7 \ M^2 \ ,
%\label{F00007}\\
\nonumber\\
m_{e^c_L}^2(M_Z) \ &\simeq& m^2_{e^c_L} + 0.23 \ M^2\ ,
\label{alfa}
\end{eqnarray}
where the second term in the expression of the scalar masses is the effect 
of gaugino loop contributions. In the above formulae we have neglected 
the scalar potential D-term contributions, which are normally small 
compared to the terms above, and the contribution of the $U(1)_Y$
D-term in the RGEs of scalar masses. 
These may be found e.g. in \cite{LLEM}. 

\subsection{Supergravity models}

We now specialize the above general discussion to the case of  
supergravity models where the observable (here MSSM) matter fields have 
diagonal metric:
\begin{eqnarray}
{\tilde K}_{\overline{\alpha} \beta}(h_m,h_m^*) = 
{\delta}_{\overline {\alpha} \beta} 
{\tilde K}_{\alpha}(h_m,h_m^*) \ .
\label{delta}
\end{eqnarray} 
This possibility is particularly interesting due to its simplicity 
and also for phenomenological reasons related to the absence of FCNC 
in the effective low-energy theory (see \cite{BIM,CEKLP,LNI,BC,BS,kiwoon} 
for a discussion on this point). Besides, the supergravity 
models that will 
be studied below correspond to this situation. Then
the K\"ahler potential (\ref{F3}), to lowest order in the 
observable fields $C^{\alpha}$, and
the superpotential (\ref{F4}) have the form 
\begin{eqnarray}
K &=& \hat K(h_m,h_m^*)+ {\tilde K}_\alpha(h_m,h_m^*) 
C^{*\overline{\alpha}} C^{\alpha} + 
\left[ Z(h_m,h_m^*) H_1 H_2 + h.c.\right]\ ,
\label{F9}\\
W &=& \hat W(h_m)+\mu(h_m) H_1 H_2 
\ + \sum_{generations} \!\! 
   \left[Y_u(h_m) Q_L H_2 u_L^c\right. 
\nonumber \\ &&
\left. +\ Y_d(h_m) Q_L H_1 d_L^c + 
Y_e(h_m) L_L H_1 e_L^c\right]\ ,
\label{F10}
\end{eqnarray}
where $ C^\alpha=Q_L, u_L^c, d_L^c, L_L,e_L^c, H_1, H_2$, 
and we have taken for simplicity diagonal Yukawa couplings
($ Y_{\alpha \beta \gamma}=Y_u,Y_d,Y_e $, in a
self-explanatory notation). 
Now the form of the effective soft Lagrangian obtained from
(\ref{F2}) and (\ref{potencial}) is given by
\begin{eqnarray}
{\cal L}_{soft} &=& \frac{1}{2}(M_a 
{\widehat{\lambda}}^a {\widehat{\lambda}}^a + h.c.)
%{\overline{{\widehat{\lambda}}}}^a {\widehat{\lambda}}^a
- m_{\alpha}^2 \widehat{C}^{*\overline {\alpha}}
\widehat{C}^{\alpha}
\nonumber\\ &&
-\
\left(\frac{1}{6} A_{\alpha \beta \gamma} \widehat{Y}_{\alpha \beta \gamma}
	    \widehat{C}^{\alpha} \widehat{C}^{\beta} \widehat{C}^{\gamma}
  + B {\widehat{\mu}} \widehat{H}_1 \widehat{H}_2+h.c.\right)\ ,
\label{F6}
\end{eqnarray}
with
\begin{eqnarray}
{m}_{\alpha}^2 &=& 
\left(m_{3/2}^2+V_0\right) - {\overline{F}}^{\overline{m}} F^n 
\partial_{\overline{m}}\partial_n \log{\tilde K_{\alpha}}\ ,
\label{mmmatrix}
\\
A_{\alpha\beta\gamma} &=& 
F^m \left[  {\hat K}_m + \partial_m \log Y_{\alpha\beta\gamma} 
- \partial_m \log({\tilde K_{\alpha}} {\tilde K_{\beta}}
{\tilde K_{\gamma}}) \right]\ ,
\label{mmmatrix2}
\\
B &=& {\widehat{\mu}}^{-1}({\tilde K}_{H_1}{\tilde K}_{H_2})^{-1/2}
\left\{ \frac{ {\hat W}^*}{|{\hat W}|} e^{{\hat K}/2} \mu 
\left( F^m \left[ {\hat K}_m + \partial_m \log\mu\right.\right.\right.
\nonumber\\ && 
\left.\left.-\ \partial_m \log({\tilde K_{H_1}}{\tilde K_{H_2}})\right]
- m_{3/2} \right)
\nonumber\\ &&
\ + 
\left( 2m_{3/2}^2+V_0 \right) {Z} -
m_{3/2} {\overline{F}}^{\overline{m}} \partial_{\overline{m}} Z
\nonumber\\ &&
+\ m_{3/2} F^m \left[ \partial_m Z - 
Z \partial_m \log({\tilde K_{H_1}}{\tilde K_{H_2}})\right]
\nonumber\\ &&
\left.-\ {\overline{F}}^{\overline{m}} F^n 
\left[ \partial_{\overline{m}} \partial_n Z - 
 \partial_{\overline{m}} Z 
\partial_n \log({\tilde K_{H_1}}{\tilde K_{H_2}})
\right] \right\}
\ ,  
\label{mmmatrix3}
\end{eqnarray}
where $\widehat{C}^{\alpha}$ and $\widehat{\lambda}^a$ are the scalar and
gaugino canonically
{\it normalized} fields respectively
\begin{eqnarray}
\widehat{C}^\alpha &=& {\tilde K}_{\alpha}^{1/2} C^\alpha\ ,
\label{yoquese}
\\
\widehat{\lambda}^a &=& (Re f_a)^{1/2} \lambda^a\ , 
\label{F7}
\end{eqnarray}
and the rescaled Yukawa couplings and $\mu$ parameter 
\begin{eqnarray}
{\widehat{Y}}_{\alpha \beta \gamma}    &=& Y_{\alpha \beta \gamma} \; 
 \frac{ {\hat W}^*}{|{\hat W}|} \; e^{{\hat K}/2} \;
({\tilde K}_\alpha {\tilde K}_\beta {\tilde K}_\gamma)^{-1/2}\ ,
\label{yyoquese}
\\
\widehat{\mu} &=& 
\left(  \frac{ {\hat W}^*}{|{\hat W}|} e^{\hat K/2} {\mu}
+ m_{3/2} Z -
{\overline {F}}^{\overline{m}} \partial_{\overline{m}} Z \right)
({\tilde K}_{H_1}{\tilde K}_{H_2})^{-1/2}\ ,
\label{rescalado}
\end{eqnarray}
have been factored out in the $A$ and $B$ terms as usual.

Now we are ready to study specific supergravity models.
%As already explained in the introduction and follows from the above 
As follows from the above
discussion, the particular values of the soft parameters 
depend on the type of supergravity theory from which the MSSM derives
and, in general, on the mechanism of SUSY 
breaking (through the presence of ${\hat W}(h_m)$ in $m_{3/2}$ and
F terms). However,
it is still possible to learn things about soft parameters without 
knowing all the details of SUSY breaking. In order to show this,
let us consider two simple and interesting supergravity models studied 
extensively in the literature: minimal supergravity and no-scale 
supergravity.

\vspace{0.1cm}

\noindent i) {\it Minimal supergravity}

\vspace{0.1cm}

\noindent This model corresponds to use the form of $K$ that leads to minimal
(canonical) kinetic terms in the supergravity 
Lagrangian, 
%\cite{Bfs,Hall}
namely  
\begin{eqnarray}
{\tilde K}_{\alpha}(h_m,h_m^*)=1 
%$K_{\alpha}(h_i,h_i^*)=1$ 
\label{minimal}
\end{eqnarray}
in (\ref{F9}). 
Then, irrespective of the SUSY-breaking mechanism, the scalar masses and
the $A$, $B$ parameters can be straightforwardly computed using
(\ref{mmmatrix}), (\ref{mmmatrix2}) and (\ref{mmmatrix3})
\begin{eqnarray}
m_{\alpha}^2	&=& m_{3/2}^2 + V_0\ , 
\label{F12} \\
A_{\alpha\beta\gamma} &=& 
F^m \left({\hat K}_m + \partial_m \log Y_{\alpha\beta\gamma} \right)\ ,
\label{F13} \\
B &=& {\widehat{\mu}}^{-1} 
\left\{  \frac{ {\hat W}^*}{|{\hat W}|} e^{{\hat K}/2}\mu
\left[ F^m \left( {\hat K}_m + \partial_m \log\mu \right)-m_{3/2} \right] 
\right.
\nonumber\\ &&
+\ \left(2m_{3/2}^2+V_0\right)Z +\
m_{3/2} \left( F^m \partial_m Z
- {\overline {F}}^{\overline{m}} \partial_{\overline{m}} Z \right)
\nonumber\\ &&
\left.\ -\ {\overline {F}}^{\overline{m}} F^n \partial_{\overline{m}}  \partial_n
Z\right\} \ ,
\label{F14}
\end{eqnarray}
where
\begin{eqnarray}
\widehat{\mu} &=& 
 \frac{ {\hat W}^*}{|{\hat W}|} e^{\hat K/2} {\mu}
+ m_{3/2} Z -
{\overline {F}}^{\overline{m}} \partial_{\overline{m}} Z
\ .
\label{rescalado2}
\end{eqnarray}
Notice that the scalar masses are automatically {\it universal} in this 
model.
%, $m_\alpha=m$. 
Further simplifications can be obtained if $Z=0$ and if the superpotential 
parameters $Y_{\alpha\beta\gamma}$ and $\mu$ do not depend
on the hidden sector fields. Under such assumptions, which are common
in the literature, (\ref{F13}) and (\ref{F14}) generate
universal $A$ parameters, as well as the 
relation
\begin{eqnarray}
B &=& A-m_{3/2} \ .
\label{F15}
\end{eqnarray}
Furthermore, if we 
assume $V_0=0$, then  $m\equiv m_{\alpha}=m_{3/2}$ and the
well known result for the $B$ parameter, $B=A-m$, is recovered.
This supergravity model is attractive for its simplicity and for the natural
explanation that it offers to the universality of the soft scalar masses.

We remark that although minimal (canonical) kinetic terms for
hidden matter, ${\hat K}(h_m,h_m^*)=\sum_m h_m h_m^*$,
are also usually assumed, we have seen that it is not 
a necessary condition in order to obtain the above results. 
Concerning the kinetic terms for vector multiplets,
it can be seen from (\ref{F2}) that the minimal (canonical)
choice $f_a=const.$ is not phenomenologically interesting, 
since it implies $M_a=0$. Nonvanishing and universal gaugino 
masses can be obtained if all the $f_a$ have the same dependence
on the hidden sector fields, i.e. $f_a(h_m)=c_a f(h_m)$ for the 
different gauge group factors of the theory. 
This is in fact what happens, at tree level, in supergravity models 
deriving from superstring theory, as we will see in the next section.
As an additional comment, we stress that relation (\ref{F15}) depends 
on the particular mechanism that is used to generate the $\mu$ parameter.
As a counter-example, notice that if one takes e.g. an $h_m$--dependent
$\mu$ as in (\ref{mu}) with $\lambda=const.$, 
instead of taking $\mu=const.$, then (\ref{F14}) gives
\begin{eqnarray}
B &=& 2 m_{3/2} + \frac{V_0}{m_{3/2}}\ , 
\label{F1512}
\end{eqnarray}
with ${\widehat{\mu}}=m_{3/2}\lambda$ from (\ref{rescalado2}). Thus the
relation (\ref{F15}) does not hold. The above result (\ref{F1512}) 
can be obtained also if one takes $\mu=0$ in the superpotential (\ref{F10})
and $Z=const.$ in the K\"ahler potential (\ref{F9}).
This also follows from our discussion above about the equivalence between 
the two mechanisms when $Z$ is an analytic function.

\vspace{0.1cm}

\noindent ii) {\it No-scale supergravity}

\vspace{0.1cm}

\noindent 
In no-scale supergravity models \cite{LN}, after the spontaneous breaking of
SUSY, 
the tree-level potential vanishes identically along some directions.
A simple example of this type of models has just one hidden field
$h$, a K\"ahler potential (\ref{F9}) with 
\begin{equation}
\hat K = -3\ \log(h+h^*) \,\, , \,\, 
{\tilde K}_\alpha = (h+h^*)^{-1}\ , 
\label{kahlerp}
\end{equation}
and a superpotential (\ref{F10}) with a hidden field independent ${\hat W}$,
i.e. ${\hat W}=const.$ 
Then, the attractive result of a
{\it vanishing} (flat) tree-level effective potential for the
hidden sector (\ref{cc}) is obtained
\begin{eqnarray}
V_0=0\ ,
\label{ccc}
\end{eqnarray}
for all
VEVs of $h$. On the other hand,
the soft parameters, using
(\ref{mmmatrix}), (\ref{mmmatrix2}) and (\ref{mmmatrix3}), are given by
\begin{eqnarray}
%M_a	&=& 0\ , 
%\label{F11} \\
m_{\alpha}^2	&=& 0
\ , 
\label{F121} \\
A_{\alpha\beta\gamma} &=& 
- m_{3/2} (h + h^*) \partial_h \log Y_{\alpha\beta\gamma} \ ,
\label{F131} \\
B &=& -{\hat \mu}^{-1} m_{3/2}(h+h^*)^2
\left\{\frac{\hat W^*}{|{\hat W}|} (h+h^*)^{-3/2}\partial_h \mu \right. 
\nonumber\\ &&
\ + \left. m_{3/2}  \left[
\partial_{h^*} Z
+\partial_{h} Z
+ (h+h^*)\partial_h \partial_{h^*}
Z\right] \right\} \ ,
\label{F141}
\end{eqnarray}
where
\begin{eqnarray}
\widehat{\mu} &=& (h+h^*)\left[\frac{\hat W^*}{|{\hat W}|} 
(h+h^*)^{-3/2}{\mu}
+ m_{3/2} Z +
m_{3/2}(h+h^*)\partial_{h^*} Z \right]
\ .
\label{rescalado3}
\end{eqnarray}
Assuming now that 
% as in the previous SUGRA model,
the $\mu$ and $Z$ coefficients and the Yukawa couplings
are 
hidden field independent,
the well known result for the soft parameters
is recovered:
\begin{eqnarray}
m_\alpha = A_{\alpha\beta\gamma} = B = 0\ . 
\label{F1555}
\end{eqnarray}
Although the above parameters are vanishing at the high scale,
gaugino masses (\ref{F2}) can induce non-vanishing values at the 
electroweak scale due to radiative corrections. 
%Besides, the attractive result of a
%{\it vanishing} (flat) tree-level effective potential for the
%hidden sector (\ref{cc}) is obtained
%
%\begin{eqnarray}
%V_0=0\ ,
%\label{ccc}
%\end{eqnarray}
%
%for all
%VEVs of $h$.

%As was showed above, the effective potential for the hidden sector is flat
%(vanishing) and therefore the VEV of the hidden scalar is undetermined at
%tree level. As a consequence the gravitino mass (\ref{gravitino})
%is also undetermined. This is the attractiveness of this model
%since radiative corrections to the effective potential may determine
%{\it dinamically} the gravitino mass to be hierarchically smaller than
%the Planck scale.

\vspace{0.2cm}

\noindent 
In conclusion, both supergravity 
models considered in this section are interesting
and give rise to concrete predictions for the soft parameters.
However, one can think of many
possible supergravity models (with different $K$, $W$ and $f$) leading to 
{\it different} results for the soft terms. This arbitrariness, as
we will see in the next section, can be ameliorated in supergravity 
models 
deriving from superstring theory, where $K$, $f$, and the 
hidden sector are more constrained.
We can already anticipate, however, that in such a context 
the kinetic terms are generically {\it not} canonical. Besides, 
although K\"ahler potentials of the no-scale type  
may appear at tree-level, the superpotentials are in general hidden field
{\it dependent}. Moreover, the Yukawa couplings $Y_{\alpha\beta\gamma}$
and the bilinear coefficients $\mu$ and $Z$  
are also generically hidden field {\it dependent}.

Finally, we remark that further constraints on the soft parameter
space of the MSSM can be obtained if one wishes to avoid low-energy charge
and color breaking minima deeper than the 
standard vacuum \cite{CLLM}. On these
grounds,
and assuming also radiative symmetry breaking with nothing but the
MSSM in between the weak scale and the Planck scale,
e.g. large regions in the parameter space 
($m$, $M$, $A$, $B$) of the minimal supergravity model i) are forbidden.
%under the  big desert  and radiative symmetry breaking assumptions.
In the limiting case $m=0$ the whole parameter space turns out to be 
excluded. This has obvious implications, e.g. for the no-scale 
supergravity model ii).
If the same kind of analysis is applied to the soft parameters 
of superstring models, again strong constraints can be obtained, 
as we will comment below.

\section{Soft terms from superstring theory}

\subsection{General parametrization of SUSY breaking}

We are going to consider N=1 four-dimensional superstrings 
where the r\^ole of hidden sector fields is effectively played by
$r$ moduli fields $T_i$, $i=1,...,r$ and the dilaton field $S$,
i.e. $h_m=S, T_i$ following the notation of the previous section.
We recall that we are denoting the $T$- and $U$-type 
(K\"ahler class and complex structure in the Calabi-Yau language)
moduli
collectively by $T_i$. The associated effective N=1 supergravity 
K\"ahler potentials (\ref{F3}), to lowest order in the matter fields, 
are of the type:
\begin{eqnarray}
K &=& {\hat K}(S,S^*,T_i,T_i^*)\
+\
{\tilde K}_{{\overline{\alpha }}{ \beta }}(T_i,T_i^*){C^*}^{\overline {\alpha}}
C^{\beta } 
\nonumber\\ &&
 +\ \left[\frac{1}{2} Z_{{\alpha }{ \beta }}(T_i,T_i^*){C}^{\alpha}
C^{\beta } + h.c. \right]\ ,
\label{kahler}
\end{eqnarray}
where at the superstring tree level
\begin{eqnarray}
{\hat K}(S,S^*,T_i,T_i^*) &=& -\log(S+S^*)\ +\ {\hat K}(T_i,T_i^*)\ .
\label{kahler2}
\end{eqnarray}
The first piece in (\ref{kahler2}) is the usual term corresponding 
to the complex dilaton $S$ that is present for any compactification.
The second piece is the K\"ahler potential of the moduli fields,
which in general depends on the compactification scheme and can
be a complicated function. For the moment we leave it generic,
but in the next subsection we will analyze some specific classes
of superstring models where it has been computed.
The same comment applies to 
${\tilde K}_{{\overline{\alpha }}{ \beta }}(T_i,T_i^*)$ and 
$Z_{{\alpha }{ \beta }}(T_i,T_i^*)$.
In the case of the superpotential (\ref{F4}), $Y_{\alpha\beta\gamma}(T_i)$
is also independent of $S$, but the non-perturbative contributions
${\hat W}(S,T_i)$ and $\mu_{\alpha\beta}(S,T_i)$ may depend in general 
on both $S$ and $T_i$. Finally, for any four-dimensional superstring
the tree-level gauge kinetic function is independent of the moduli sector
and is simply given by
\begin{eqnarray}
{f_a} &=& k_a S\ , 
\label{kahler3}
\end{eqnarray}
where $k_a$ is the Kac--Moody level of the gauge factor.
Usually (level one case) one takes $k_3=k_2=\frac{3}{5}k_1=1$, but this is
irrelevant for our tree-level computation since $k_a$ will not contribute to
the soft parameters.

As we will show below, it is important to know what fields, either $S$ 
or $T_i$, play the predominant
role in the process of SUSY breaking. This will have relevant consequences
in determining the pattern of soft parameters, and therefore the spectrum
of physical particles \cite{BIM}. That is why it is very useful to introduce the
following parametrization, consistent with (\ref{cc}), 
for the VEVs 
of dilaton and moduli auxiliary fields
\begin{eqnarray}
F^S &=& \sqrt{3}Cm_{3/2}K_{{\overline S}S}^{-1/2}\sin\theta 
e^{-i\gamma _S}\ \ , 
\nonumber \\
F^i &=& \sqrt{3}Cm_{3/2}\cos\theta P^{i{\overline j}} \Theta_{\overline j}
\ ,  
\label{auxi}
\end{eqnarray}
where the constant $C$ is defined as follows
\begin{eqnarray}
C^2 &=& 1+\frac{V_0}{3m_{3/2}^2} \ .
\label{cos}
\end{eqnarray}
This parametrization is valid for the general case of off-diagonal
moduli metric, since $P$ is a matrix canonically normalizing the moduli
fields, i.e. $P^{\dagger}\hat K P=1$ where 
$\hat K \equiv {\hat K}_{{\overline i}j}$ and 1 stands for the unit matrix. 
%$P$ can be written as $P=U{\hat K}_d^{-1/2}$, where $U$ is a unitary matrix
%which diagonalizes $\hat K$ to ${\hat K}_d$.
The angle $\theta$ and the complex parameters 
$\Theta_{\overline j}$ just parametrize the direction of the goldstino
in the $S$, $T_i$ field space (see below (\ref{auxiliary})) 
and $\sum_j \Theta_j^*\Theta_{\overline j}=1$.
We have also allowed for the possibility of some complex phases which
could be relevant for the CP structure of the theory
(see \cite{BIM,CH,CEKLP,LNI,BS,BKL} 
for a discussion on this point). Notice that if the
tree-level cosmological constant $V_0$ is assumed to vanish, one has
$C=1$, but we prefer for the moment to leave it undetermined as we did
in the previous section (see below (\ref{cc})). 

Notice that such a phenomenological approach allows us
to `reabsorb' (or circumvent) our ignorance about the (nonperturbative)
$S$- and $T_i$- dependent part of the superpotential (\ref{F4}), 
${\hat W(S,T_i)}$, which is
responsible for SUSY breaking.

It is now a straightforward
exercise,
plugging (\ref{kahler}), (\ref{kahler2}), (\ref{kahler3}) and 
(\ref{auxi}) into (\ref{F2}), (\ref{mmatrix}), (\ref{mmatrix2}) and 
(\ref{mmatrix3}),
to compute the soft SUSY-breaking parameters as functions of $\theta$ and
$\Theta_{\overline j}$.  
On the one hand, 
since the tree-level gauge kinetic function is given for any 
four-dimensional superstring by (\ref{kahler3}),
the tree-level gaugino masses are universal, independent of the
moduli sector, and simply given by:
\begin{eqnarray}
M_a &=& \sqrt{3}Cm_{3/2} \sin\theta e^{-i\gamma _S} \ . 
\label{gaugin}
\end{eqnarray}
On the other hand, the bosonic soft parameters depend in general on 
the moduli sector (i.e. on the functions 
${\tilde K}_{{\overline{\alpha }}{ \beta }},
Z_{{\alpha }{ \beta }}(T_i,T_i^*),\ldots$ and on the parameters
$\cos\theta$ and $\Theta_{\overline j}$)
and therefore they should be studied in the context of specific classes of 
superstring models. However, we will first focus on the very interesting 
limit $\cos\theta=0$, which corresponds to the case where the dilaton sector is
the source of all the SUSY breaking (see (\ref{auxi})) and the results 
are compactification independent. 

\vspace{0.1cm}

\noindent {\it Dilaton SUSY breaking}

\vspace{0.1cm}

\noindent 
Since the dilaton couples in a universal manner to all particles, 
this limit is quite model {\it independent} \cite{KL,BIM}. 
Indeed, the expressions
for all the soft parameters (except $B$) are quite simple and 
independent of the four-dimensional superstring considered. After canonically 
normalizing the fields, one obtains:
\begin{eqnarray}
m_{\alpha}^2 &=& m_{3/2}^2\ +\ V_0 \ , 
\label{uno}\\
M_a &=& 
 \ \sqrt{3}C m_{3/2}e^{-i\gamma_S}  \ ,
\label{dos}\\
A_{\alpha \beta \gamma } &=&
- M_a \ ,
\label{tres}\\ 
B &=& {\widehat{\mu}}^{-1}({\tilde K}_{H_1}{\tilde K}_{H_2})^{-1/2}
\left\{ \frac{ {\hat W}^*}{|{\hat W}|} e^{{\hat K}/2}
\mu m_{3/2}\left( -1\right.\right.
\nonumber\\ &&
\left.\left.   -\ \sqrt{3}Ce^{-i\gamma_S}
\left[1-\left(S+S^*\right)\partial_S \log\mu\right]\right) 
%\right.
%\nonumber\\ &&
+Z\left( 2m_{3/2}^2+V_0\right) 
\right\}
\ ,  
\label{dilaton}
\end{eqnarray}
where
\begin{eqnarray}
\widehat{\mu} &=& 
\left(  \frac{ {\hat W}^*}{|{\hat W}|} e^{\hat K/2} {\mu}
+ m_{3/2} Z \right) ({\tilde K}_{H_1}{\tilde K}_{H_2})^{-1/2} \ .
\label{dilaton2}
\end{eqnarray}
Although the general expression for $B$ is more involved than the 
ones of the other soft parameters, a considerable simplification occurs
if $Z$ is the only source of the $\mu$ term. In this case $B$ 
reduces to  
\begin{eqnarray}
B &=& 2m_{3/2}\ +\ \frac{V_0}{m_{3/2}}\ .
\label{dilaton3}
\end{eqnarray}
and thus becomes independent of the four-dimensional superstring considered,
as the other parameters.
It is easy to check that the same result (\ref{dilaton3}) is also obtained 
if $Z=0$ and the superpotential contains a $\mu$ coefficient of the 
form (\ref{mu}), where now $\mu=\lambda(T_i){\hat W}(S,T_i)$.  
Notice that the expressions for
$m_{\alpha}$ (\ref{uno}) and $B$ (\ref{dilaton3}) coincide with the 
corresponding ones obtained in the minimal supergravity 
model i), (\ref{F12}) and 
(\ref{F1512}) respectively. Furthermore, if $Z=0$ and $\partial_S\mu=0$,
the expression for $B$ obtained from (\ref{dilaton}) coincides 
with the corresponding one (\ref{F15}) of the minimal supergravity model.

This dilaton-dominated scenario 
is attractive for its simplicity and for
the natural explanation that it offers to the {\it universality} of the
soft terms. 
For possible explicit SUSY--breaking mechanisms where this limit might
be obtained see \cite{macorra}. 
%Actually, universality is a desirable property not
%only to reduce the number of independent parameters in the MSSM, but also
%for phenomenological reasons, particularly to avoid FCNC. 
Because of the simplicity of this scenario, the low-energy predictions 
are quite precise \cite{BLM,BIM,LNZ,KMV}. Assuming 
a vanishing cosmological constant and imposing,
e.g. from the limits on the electric dipole moment of the neutron,
$\gamma_S$ = $0$ mod $\pi$ 
(\ref{uno}), (\ref{dos}) and (\ref{tres}) give 
\footnote{
It is worth remarking that  these particular boundary conditions
have also interesting finiteness properties. In particular, they preserve
one-loop finiteness  of $N=1$ finite theories \cite{strings95}.} 
%See e.g. \cite{strings95} and references
%therein.}
%
\begin{eqnarray}
& m_{\alpha}\ =\ m_{3/2}\ , \ M_a\ =\ \pm \sqrt{3} m_{3/2}\ ,\
A_{\alpha\beta\gamma}\ =\ -M_a\ . &
\label{cuatro}
\end{eqnarray}
Since scalars are lighter than gauginos at the high scale
and all mass ratios are fixed, at low-energy ($\sim M_Z$) one
finds the following mass ratios for the gluino, slepton and
squark (except stops and left sbottom) masses 
\begin{eqnarray}
 M_g:m_{Q_L}:m_{u_L^c}:m_{d_L^c}:m_{L_L}:m_{e_L^c}  
\simeq 1:0.94:0.92:0.92:0.32:0.24 \ , 
\label{dilaton4}
\end{eqnarray}
as can be computed e.g. from (\ref{alfa}). 
%(actually, here we have 
%used slightly modified numerical factors since we have taken
%the superstring scale as boundary scale). 
Although squarks and sleptons have the same soft mass at the high
scale, at low-energy the former are much heavier than the latter 
because of the gluino contribution to the renormalization of their masses.
The rest of the spectrum is very dependent on the details of 
$SU(2)_L\times U(1)_Y$ breaking and therefore on the values of 
$B$ and ${\widehat{\mu}}$.
For $B=2m_{3/2}$ (see(\ref{dilaton3})) and $\mu$ treated as a free parameter
this analysis can be found in
\cite{BLM}. Modifications to this scenario due to the effect of possible
superstring non--perturbative corrections to the K\"ahler potential can
be found in \cite{albertus}.

It is worth noticing here that, although the value of 
${\widehat{\mu}}$ (\ref{dilaton2}) is compactification dependent 
even in this dilaton-dominated scenario, the simple result 
${\widehat{\mu}}=m_{3/2}$ can be obtained in any compactification
scheme where the source of ${\widehat{\mu}}$ is a $Z$ term 
in the K\"ahler potential fulfilling the property
$Z=({\tilde K}_{H_1}{\tilde K}_{H_2})^{1/2}$. 
In fact, we will see in the next subsection that this is the case 
of some classes of orbifold models.
Notice that, when such a property holds, the whole SUSY spectrum 
depends only on one parameter, $m_{3/2}$, since
\begin{eqnarray}
m_{\alpha }=m_{3/2} \ , 
\ M_a =
\pm \sqrt{3} m_{3/2}   \ , 
\ A_{\alpha \beta \gamma } =
- M_a \ , 
\ B = 2 m_{3/2} \ , 
\ \widehat{\mu} =  m_{3/2} \ .
\label{dilaton8}
\end{eqnarray} 
%If we would know the particular mechanism which breaks SUSY, then we would
%be able of computing the superpotential and hence 
%$m_{3/2}=e^{\hat K/2}|\hat W|$. 
%Although this is not the case,
Besides, this parameter can be fixed from the phenomenological requirement
of correct electroweak breaking $2M^2_W/g_2^2=\langle |H_1|^2 \rangle
+\langle |H_2|^2 \rangle$. Thus at the
end of the day we are left essentially with no free parameters. 
%Of course, if in the next future the mechanism
%which breaks SUSY is known (i.e. $m_{3/2}$ can be explicitly calculated) 
%and the above scenario is the correct one, the
%value of $m_{3/2}$ should coincide 
%with the one obtained from the phenomenological
%constraint.
In \cite{BIM2} the consistency of the above boundary conditions with the
appropriate radiative electroweak symmetry breaking was explored. 
Unfortunately, it was 
found that they are not consistent with the measured value of the top-quark
mass, namely the mass obtained in this scheme turns out to be too small.
A possible way-out to this situation is to assume that also the moduli
fields contribute to SUSY breaking, since the soft terms are then
modified. Of course, this amounts to a departure of the pure 
dilaton-dominated scenario. This possibility will be discussed in 
the context of orbifold models in the next subsection.

Finally, we recall that the phenomenological problem of the pure 
dilaton-dominated limit mentioned above is also obtained in a different
context, namely from requiring the absence of low-energy charge and 
color breaking minima deeper than the standard vacuum \cite{CLLM2}. 
In fact, on these grounds, the 
dilaton-dominated limit is excluded not only for a $\mu$ term generated
through the K\"ahler potential but for any possible mechanism solving the
$\mu$ problem. The results indicate that the whole free parameter space
($m_{3/2}$, $B$, $\mu$) 
is excluded after imposing the present experimental data
on the top mass. 
Again this rests on the assumption of
radiative symmetry breaking with nothing but the MSSM in between the
weak scale and the superstring scale.
%The inclusion of a non-vanishing tree-level 
%cosmological constant
%does not improve esentially this situation.

\vspace{0.1cm}

\noindent {\it Dilaton/Moduli SUSY breaking}

\vspace{0.1cm}

\noindent
In general the moduli fields $T_i$ may also contribute to SUSY breaking,
i.e. $F^i\neq 0$ in (\ref{auxi}),
and therefore their effects on soft parameters must also be 
included \cite{BIM,FKZ,KSYY,BC,BIMS}. In this sense it is interesting to
note that explicit possible scenarios of SUSY breaking by gaugino
condensation in superstrings, when analyzed at the one--loop level, 
lead to the mandatory inclusion of the moduli in the game (in fact the
moduli are the main source of SUSY breaking in these cases) \cite{gaugino}.
Since different compactification schemes give rise to different
expressions for the moduli-dependent part of the K\"ahler potential
(\ref{kahler}), the computation of the bosonic soft parameters will be 
model {\it dependent}. The results are discussed below in the context of
some specific superstring models.

\subsection{Superstring models}

To illustrate the main features of mixed dilaton/moduli SUSY breaking,
we will concentrate mainly on the case of diagonal moduli and matter 
metrics. For instance, under this assumption the parametrization (\ref{auxi})
is simplified to
\begin{eqnarray}
F^S &=& \sqrt{3}Cm_{3/2}{\hat K}_{{\overline S}S}^{-1/2}\sin\theta 
e^{-i\gamma _S}\ \ , 
\nonumber \\
F^i &=& \sqrt{3}Cm_{3/2}{\hat K}_{{\overline i}i}^{-1/2}\cos\theta \Theta_{i}
e^{-i\gamma_i}
\ , 
\label{auxili}
\end{eqnarray}
where $\sum_i \Theta_i^2=1$.
Although this is the generic case e.g. in most orbifolds, off--diagonal
metrics are present in general in Calabi--Yau compactifications. This
may lead to FCNC effects in the low--energy effective N=1 softly broken
Lagrangian. The analysis of soft SUSY-breaking parameters in Calabi--Yau
compactifications is therefore more involved and can be found 
in \cite{KM}
using parametrization (\ref{auxi}).
A similar analysis for the few orbifolds with off--diagonal metrics
was carried out in \cite{BIMS}. 
Some comments about the ``off-diagonal" 
results will be made below. 
Also in the case of orbifold compactifications with continuous Wilson 
lines off-diagonal moduli metrics arise, due to the moduli--Wilson line 
mixing. However, this analysis turns out to be simple \cite{KM2} and 
the results are similar to the ones studied below in the diagonal case. 
%In any case several comments about the 
%off--diagonal orbifold results will be made below.

Since the moduli part of the K\"ahler potential (\ref{kahler})
has been computed for $(0,2)$ symmetric Abelian orbifolds, 
we will concentrate here on these models.
They contain  generically
three $T$-type moduli 
(the exceptions are the orbifolds $Z_3$, 
$Z_4$ and $Z'_6$, which have 9, 5 and 5 respectively, and
are precisely the ones with off-diagonal
metrics) 
and, at most, three $U$-type moduli.
We will denote them collectively by $T_i$, 
where e.g. $T_i=U_{i-3}$; $i=4,5,6$.
For this  class of models the K\"ahler potential has the 
form 
\begin{eqnarray}
K &=& -\log(S+S^*) - \sum _i \log(T_i+T_i^*) 
+ \sum _{\alpha } |C^{\alpha }|^2 \Pi_i(T_i+T_i^*)^{n_{\alpha }^i} \ .
\label{orbi}
\end{eqnarray}
Here $n_{\alpha }^i$ are (zero or negative) fractional numbers usually 
called ``modular weights" of the matter fields $C^{\alpha }$. 
For each given Abelian orbifold,
independently of the gauge group or particle content, the possible
values of the modular weights are very restricted. For a classification of
modular weights for all Abelian orbifolds see \cite{IL}.
%As a matter of fact, the K\"ahler potentials which appear in the large-$T$
%limit of Calabi-Yau compactifications \cite{calabi} and 
%4-D fermionic Strings \cite{fermionic}
%are quite close to the above one. Thus the results that we will
%obtain below will probably be more general than just for orbifold 
%compactifications.
The piece proportional to $Z_{\alpha\beta}$ in (\ref{kahler})
has been shown to be present in Calabi--Yau compactifications and 
orbifolds. In particular, in the case of orbifolds, such a term
arises when the untwisted sector has at least one complex--structure 
field $U$ and has been explicitly computed.
We will analyze separately this case below, as well as the associated 
$\mu$ and $B$ parameters, whereas we will concentrate 
here on the other bosonic soft parameters. Plugging 
the particular form (\ref{orbi}) of the K\"ahler potential and
the parametrization (\ref{auxili}) 
in (\ref{mmmatrix}) and (\ref{mmmatrix2}) 
we obtain
the following results
%\footnote{This analysis was also carried out for the
%particular case of the three diagonal moduli $T_i$
%in ref.\cite{japoneses} and \cite{BC}
%in order to obtain unification of gauge coupling constants
%and to analyze  
%FCNC constraints respectively.
%Some particular multimoduli examples were also considered in
%ref.\cite{FKZ}.} 
for the scalar masses and trilinear parameters \cite{BIMS,KSYY,BC}:
\begin{eqnarray}
m_{\alpha }^2 &=& m_{3/2}^2\left(1 + 3C^2\cos^2\theta\ {\vec {n_{\alpha }}}.
{\vec {\Theta ^2}}\right)\ +\ V_0\ ,
\label{masorbi2}\\
%&  M = \  \sqrt{3}m_{3/2}\sin\theta e^{-i{\gamma }_S} \ , &
%\nonumber\\
A_{\alpha \beta \gamma } &=& -\sqrt{3} Cm_{3/2} \left( \sin\theta e^{-i{\gamma
}_S}
%\right.
%\nonumber\\ &&
+ \cos\theta \sum _{i=1}^6 e^{-i\gamma _i}    {\Theta }_i 
\left[1\right.\right.
\nonumber\\ &&
\left.\left. +\
n^i_{\alpha }+n^i_{\beta
}+n^i_{\gamma
}-
(T_i+T_i^*) \partial_i \log Y_{\alpha \beta \gamma}\! \right]\right)\ . 
\label{masorbi}
\end{eqnarray}
It is easy to check that the results (\ref{uno}) and (\ref{tres}) 
are recovered in the limit where $\cos\theta \rightarrow 0$.
Notice that neither the scalar (\ref{masorbi2}) nor the gaugino masses 
(\ref{gaugin}) have any explicit dependence on $S$ or $T_i$: they 
only depend on the gravitino mass and the goldstino angles.
This is one of the advantages of a parametrization in terms of such angles.
Although in the case of the $A$-parameter 
an explicit $T_i$-dependence may appear in
the term proportional to $\partial_i \log Y_{\alpha \beta \gamma }$, 
it disappears in 
several interesting 
cases \cite{BIMS}. 
Using the above information, we can now analyze the structure of
soft parameters available in Abelian orbifolds.

In the dilaton-dominated case ($\cos\theta =0$) the 
soft parameters are universal, as already studied in the previous section.
However, in general, they show a lack of universality due to the
modular weight dependence (see (\ref{masorbi2}) and (\ref{masorbi})).
So, even with diagonal matter metrics, FCNC effects may appear. However,
we recall that the low-energy running of the scalar masses has to be
taken into account. In particular, in the squark case, for gluino masses 
heavier than (or of the same order as) the scalar masses at the boundary
scale, there are large flavour-independent gluino loop contributions 
which are the dominant source of scalar masses (see (\ref{alfa})).
We will show below that this situation is very common in orbifold models.
The above effect can therefore help in fulfilling the FCNC constraints.

Another feature of the case under study is that, depending on the 
goldstino direction, tachyons may appear. For $\cos^2\theta \geq 1/3 $, 
the goldstino direction cannot be chosen arbitrarily if one is interested
in avoiding tachyons (see (\ref{masorbi2})). 
Nevertheless, having a tachyonic sector is
not necessarily a problem, it may even be an advantage \cite{BIMS}.
In the case of superstring GUTs 
(or the standard model with extra U(1) interactions),
the negative squared mass may just induce gauge symmetry breaking by 
forcing a VEV 
for a particular scalar, GUT-Higgs field, in the model.
The latter possibility provides us with interesting phenomenological
consequences: the breaking of SUSY could directly induce further 
gauge symmetry breaking.

Finally, let us consider three particles
$C^{\alpha }$, $C^{\beta }$ and $C^{\gamma }$, 
coupled through a Yukawa $Y_{\alpha \beta \gamma }$. They may belong
both
to the untwisted (${\bf U}$) sector or to a twisted 
(${\bf T}$) sector, i.e. we consider couplings
of the type ${\bf U}{\bf U}{\bf U}$, 
${\bf U}{\bf T}{\bf T}$,
${\bf T}{\bf T}{\bf T}$. Then, using the above formulae (\ref{masorbi2}) and
(\ref{gaugin}), with negligible $V_0$,
%and phases for simplicity, 
one
finds \cite{BIMS}
that in general for {\it any} choice of goldstino 
direction
%\footnote{It is worth noticing that a similar sum rule can be
%found in gauge-Yukawa unified models. For an analysis of this 
%`coincidence' see \cite{coincidencia}.}
%
\begin{equation}
m_{\alpha }^2\ +\ m_{\beta }^2\ +\ m_{\gamma }^2\ \leq \ |M_a|^2\
=\ 3 m_{3/2}^2\sin^2\theta \ .
\label{rulox}
\end{equation}
Remarkably, the same sum rule is fulfilled even in the presence of
off-diagonal metrics, as it is the case of the orbifolds $Z_3$, $Z_4$ and
$Z'_6$. The three scalar mass eigenvalues will be in general non-degenerate, 
which in turn may induce FCNC. This can be automatically avoided in the
dilaton dominated limit or under special conditions (for instance, when 
$\hat W$ does not depend on the moduli, a no-scale scenario arises and
the mass eigenvalues vanish). The same problem is present 
in Calabi-Yau compactifications, where again the mass eigenvalues are 
typically non-degenerate. Besides, the sum rule (\ref{rulox}) is violated 
in general \cite{KM}. Coming back to the orbifold case, notice that
the above sum rule implies that on average scalars are lighter than 
gauginos. For small $\sin\theta$, some particular scalar mass may
become bigger than the gaugino mass, but in that case at least one
of the other scalars involved in the sum rule would be forced
to have a {\it negative} squared mass. This situation is quite dangerous 
in the context of standard model four-dimensional superstrings, 
since some observable particles, 
like Higgses, squarks or sleptons, could be forced to acquire large VEVs 
(of order the superstring scale). 
%For example, the scalars
%associated through the Yukawa coupling $H_2Q_Lu_L^c$, which generates the
%mass of the $u$-quark, must fulfil the above sum-rule
%(\ref{rulox}). If we
%allow e.g. the scalars $H_2$, $Q_L$ to be heavier than gauginos, then
%$u_L^c$ will become tachyonic breaking charge and color.
%
%However, tachyons may be helpful if the particular Yukawa coupling
%does not involve observable particles. They could break extra gauge symmetries
%and generate large masses for extra particles. We recall that standard-like
%models in Strings usually have too many extra particles and many extra
%U(1) interactions. Although the Fayet-Iliopoulos mechanism helps to cure
%the problem \cite{suplemento}, the existence of tachyons is a complementary
%solution.
%We thus see that, for standard model superstrings,
%if we want  to avoid charge and colour breaking minima (or VEVs of
%order the string scale for the 
%Higgses\footnote{For a possible way-out to this problem, allowing the 
%possibility of scalars heavier than gauginos, see ref.\cite{nuevo}.}),
%we should grosso modo come back to a situation
%with gauginos heavier than scalars.
%, as in ref.\cite{BIM}.
If the above sum rule is applied and squared soft masses are 
(conservatively) required to be non-negative in order to avoid 
instabilities of the scalar potential, then the tree level soft 
masses of observable scalars are constrained to be always smaller 
than gaugino masses at the boundary scale:
\begin{eqnarray}
m_{\alpha} &<& M_a\ .
\label{vaya}
\end{eqnarray}
In turn, this implies that, at low-energy ($\sim M_Z$), the masses
of gluinos, sleptons and squarks (except stops and left sbottom)
are ordered as 
\begin{eqnarray}
& m_l\ <\ m_q\ \simeq M_g \ , &
\label{masas2}
\end{eqnarray}
where gluinos are slightly heavier than scalars. Therefore,
in spite of the different set of (non-universal) soft scalar masses,
the low-energy phenomenological predictions of the mixed dilaton/moduli 
SUSY breaking become qualitatively similar to those of the pure 
dilaton-dominated SUSY breaking. This holds especially for the 
squark masses, as follows e.g. from (\ref{vaya}) and 
(\ref{alfa}). In the case of sleptons, which do not feel gluino
loop effects, the boundary values of the soft masses (\ref{vaya}) 
can be relatively 
more important at low-energy, and larger deviations 
from the numerical results of (\ref{dilaton4}) can be obtained.
Analyses of the low-energy predictions of the dilaton/moduli scenario
taking account of the radiative symmetry breaking 
can be found
in \cite{BIM,KMV,CDG}.

Before concluding, we recall that exceptions to the above pattern 
(\ref{vaya}), (\ref{masas2}) can arise in several situations \cite{BIMS}. 
For instance, since the total squared Higgs masses receive a positive 
contribution $\mu^2$, the corresponding soft masses may be allowed 
to be negative: in this case the restrictions from the sum rule would 
be relaxed.  Another example concerns MSSM Yukawa couplings that arise 
effectively from higher dimension operators: in this case the three-particle 
sum rule itself may not hold. Finally, a departure from relations 
(\ref{vaya}) and (\ref{masas2}) can also arise when both scalar and gaugino 
masses vanish at tree level. Such a vanishing can happen in the 
fully moduli-dominated SUSY breaking, e.g. if SUSY breaking is equally 
shared among $T_1, T_2, T_3$ and one consider untwisted particles: 
%(for an analysis of the low-energy predictions see 
%\cite{BIM,KMV,CDG}):
then superstring loop effects become important and tend to make scalars
heavier than gauginos \cite{BIM}. In any event, we stress again that potential
violations of the pattern in (\ref{vaya}) and (\ref{masas2}) can occur 
only when SUSY breaking is mainly moduli dominated (specifically,
$\cos^2\theta \geq 2/3$), since only in this case the gaugino masses 
can decrease below $m_{3/2}$ and possibly become lighter than some 
scalar mass.

\vspace{0.1cm}

\noindent 
{\it The $B$ parameter and the $\mu$ problem}

\vspace{0.1cm}

\noindent 
As already discussed in section 2.1, the two mechanisms
to solve the $\mu$ problem in the context of supergravity are {\it naturally}
present in superstring models. We will concentrate here on the case
in which $\mu$ arises from a bilinear term in the K\"ahler 
potential (\ref{F3}). The alternative mechanism which generates
$\mu$ from the superpotential, as in (\ref{mu}), may also be
present in orbifolds, but the results are more model dependent. They
can be found in \cite{BIMS}. 
We recall that, in any orbifold with at least one complex-structure 
field $U$, the K\"ahler potential of the untwisted sector possesses
the structure $Z(T_i,T_i^*)C_1C_2+h.c.$ \cite{LLM,AGNT} and can therefore
generate a $\mu$ term. Specifically, the $Z_N$ orbifolds
based on $Z_4$, $Z_6$, $Z_8$, $Z_{12}'$ and the $Z_N\times Z_M$ orbifolds based
on $Z_2\times Z_4$ and $Z_2\times Z_6$ do all have a $U$-type field in (say)
the third complex plane. In addition, the $Z_2\times Z_2$ orbifold has $U$
fields in the three complex planes.
In all these models the piece of the K\"ahler potential involving
the moduli and the untwisted matter fields $C_{1,2}$ in the third complex
plane has the form
\begin{eqnarray}
K_3
%(T_i,T_i^*,C_1,C_2) 
&=& 
-\log\left[(T_3+T_3^*)(U_3+U_3^*) - (C_1+C_2^*)
(C_1^*+C_2)\right]
\label{kahlb} \\
&\simeq&
 - \log(T_3+T_3^*)  - \log(U_3+U_3^*)\ +
\frac{(C_1+C_2^*)(C_1^*+C_2)}{(T_3+T_3^*)(U_3+U_3^*)} \ .
\label{kahlexp}
\end{eqnarray}
Now, from the expansion shown in (\ref{kahlexp}), one can easily read off the
functions $Z$, ${\tilde K}_1$, ${\tilde K}_2$ associated to $C_1$ and $C_2$:
\begin{equation}
Z\ =\ {\tilde K}_1 \ =\ {\tilde K}_2\ =\  {1\over {(T_3+T_3^*)(U_3+U_3^*)}}
\ .
\label{zzz}
\end{equation}
Let us assume that the MSSM can be obtained from a superstring model 
of the kind mentioned above and let us identify the
fields $C_1$ and $C_2$ with the electroweak Higgs fields $H_1$ and $H_2$.
Plugging back the expressions (\ref{zzz}) in (\ref{mmmatrix3}) 
and (\ref{rescalado}) with $\mu=0$, and using
the parametrization (\ref{auxili}),
one can compute $\mu$ and $B$ for this interesting class
of models \cite{BIMS}:
\begin{eqnarray}
\widehat{\mu} &=& m_{3/2} \left[1 + \sqrt{3}C\cos\theta
(e^{i \gamma_3} \Theta _3 + e^{i \gamma_6} \Theta _6)\right] \ , 
\label{muu}
\\
B\widehat{\mu} &=& 2m_{3/2}^2 \left[1 +\sqrt{3}C \cos\theta
 ( \cos\gamma_3 \Theta_3 + \cos\gamma_6 \Theta_6) \right. 
\nonumber\\ &&
\left. +\ 3C^2\cos^2\theta \cos(\gamma_3-\gamma_6) {\Theta _3}{\Theta _6} 
\right]\ +\ V_0\ . 
\label{bmu}
\end{eqnarray}
Notice that, in the limit where $\cos\theta \rightarrow 0$,
the results in (\ref{dilaton8}) are recovered.
In addition, we recall from (\ref{masorbi2}) that the soft masses are
\begin{eqnarray}
& m^2_{H_1}\ =\ m^2_{H_2}\ =\  m_{3/2}^2\ \left[ 1\ -\ 3C^2\cos^2\theta
(\Theta_3^2+\Theta _6^2)\right]\ +\ V_0\ . &
\label{mundos}
\end{eqnarray}
In general, the quadratic part of the Higgs potential 
after SUSY breaking has
the form (see(\ref{F6}))
\begin{eqnarray}
V_2\ =\ (m_{H_1}^2+|\widehat{\mu}|^2)|\widehat{H}_1|^2\ 
+\ (m_{H_2}^2+|\widehat{\mu}| ^2)|\widehat{H}_2|^2\  +\
(B\widehat{\mu} \widehat{H}_1\widehat{H}_2+h.c.)\ ,
\label{flaty}
\end{eqnarray}
where we recall that ${\hat H}_1$ and ${\hat H}_2$ are the
canonically normalized Higgs fields.
In the specific case under consideration, 
from
(\ref{muu}), (\ref{bmu}) and (\ref{mundos}) we find the remarkable result
that the three coefficients in $V_2$ are equal, i.e.
\begin{eqnarray}
& m_{H_1}^2\ +\ |\widehat{\mu}|^2\ =\ m_{H_2}^2\ +\ |\widehat{\mu}| ^2\ =
\ B\widehat{\mu} \ . &
\label{result}
\end{eqnarray}
so that $V_2$ has the simple form
\begin{eqnarray}
V_2 &=&  B\widehat{\mu} \ (\widehat{H}_1+\widehat{H}_2^*)
(\widehat{H}_1^*+\widehat{H}_2) \ . 
\label{potflat}
\end{eqnarray}
and therefore $\tan\beta=\frac{<{\hat H}_2>}{<{\hat H}_1>}=-1$.
Of course, this corresponds to the boundary condition on the
scalar potential at the superstring scale: at lower energies
the renormalization group equations should be used.
Although the common value of the three coefficients in (\ref{result})
depends on the Goldstino direction via the parameters
$\cos\theta$, $\Theta_3$, $\Theta_6$,\ldots (see e.g. the
expression of $B\widehat{\mu}$
in (\ref{bmu})), we stress that the equality itself and the form
of $V_2$ hold {\em independently} of the Goldstino direction.

Starting from such `superstringy' boundary conditions for the MSSM
parameters, one can explore their consistency with radiative
electroweak-symmetry breaking \cite{BIM2} (see also \cite{KKW}). 
One finds that consistency with  
the measured value of the top-quark mass cannot be achieved 
in the dilaton-dominated scenario (as already mentioned in
section 3.1). 
The only SUSY-breaking scenario 
compatible with such constraints requires a suppressed dilaton
contribution and important (often dominant) contributions from the
$T_3$, $U_3$ moduli.

\section{Final Comments and Outlook}

It is worth remarking that most  of  the above results for soft terms 
in superstring models refer to certain simple {\it perturbative} 
heterotic compactifications.  In addition,  it is assumed that the 
goldstino  is  a fermionic partner of some combination of the
dilaton and/or the moduli  fields. Recently some information about
the non-perturbative regime in superstring theory has been obtained
in terms of the S-dualities \cite{filq} 
of the theory.  All superstring theories  seem to 
correspond to some points in the parameter space of a unique
eleven-dimensional underlying theory, M-theory \cite{mtheory}. 
Although the structure of this
theory is largely unknown,  some preliminary attempts have been made 
to extract some information of phenomenological interest.
A scenario to understand the difference  between the GUT scale and
the superstring scale 
has been put forward \cite{W}.
Supersymmetry breaking and other phenomenological issues   
have also been explored within this
context in \cite{mtpheno} .
Studies in progress concerning {\it non-perturbative } 
superstring vacua with $N=1$ SUSY will certainly  bring us
new surprises. 
%\section*{Acknowledgements}
%
%Gracias a la viiiidaaaaa  ,  que me ha dado taaaaantooooo......


\centerline{\bf References}

\begin{thebibliography}{99}
%
\bibitem{HPN} For a review, see: H.P. Nilles, {\it Phys. Rep.}
{\bf 110} (1984) 1, and references therein.
%
\bibitem{CM} For a recent review, see: C. Mu\~noz, 
%`Soft supersymmetry--breaking terms and the $\mu$ problem',
%{\it FTUAM} {\bf 95/20},
{\it hep-th/9507108}, and references therein.
%
\bibitem{CM2} For a review, see: C. Mu\~noz, 
%`Soft terms from strings', 
%{\it FTUAM} {\bf 96/4},
{\it hep-ph/9601325}, and references therein.
%
\bibitem{KKK} Y. Kawamura, T. Kobayashi and T. Komatsu,
%{\it DPSU-96-12},
{\it hep-ph/9609462}, and references therein.
%
\bibitem{SW} S.K. Soni and H.A. Weldon, 
{\it Phys. Lett.} {\bf B126} (1983) 
215.
%
\bibitem{BIM} A. Brignole, L.E. Ib\'{a}\~{n}ez and C. Mu\~noz,
{\it Nucl. Phys.} {\bf B422} (1994) 125 [Erratum: {\bf B436} (1995) 747].
%
\bibitem{CKN} K. Choi, J.E. Kim and H.P. Nilles, 
{\it Phys. Rev. Lett.} {\bf 73} (1994) 1758;
K. Choi, J.E. Kim and G.T. Park,
{\it Nucl. Phys.} {\bf B442} (1995) 3.
%
\bibitem{FKZ} S. Ferrara, C. Kounnas and F. Zwirner,
{\it Nucl. Phys.} {\bf B429} (1994) 589 [Erratum: {\bf B433} (1995) 255].
%
\bibitem{kiwoon} K. Choi, J.S. Lee and C. Mu\~noz, {\it hep-ph/9709250}.
%
\bibitem{GM} G.F. Giudice and A. Masiero, {\it Phys. Lett.} {\bf B206}
(1988) 480.
%
\bibitem{JACM} J.A. Casas and C. Mu\~noz, {\it Phys. Lett.} {\bf B306}
(1993) 288.
%
\bibitem{KN} 
J.E. Kim and H.P. Nilles, 
{\it Phys. Lett.} {\bf B138} (1984) 150,
{\it Phys. Lett.} {\bf B263} (1991) 79;
E.J. Chun, J.E. Kim and H.P. Nilles,
{\it Nucl. Phys.} {\bf B370} (1992) 105.
%
\bibitem{KL} V.S. Kaplunovsky and J. Louis 
{\it Phys. Lett.} {\bf B306} (1993) 269.
%
\bibitem{LLM} G. Lopes-Cardoso, D. L\"ust and T. Mohaupt,
{\it Nucl. Phys.} {\bf B432} (1994) 68. 
%
\bibitem{AGNT}
I. Antoniadis, E. Gava, K.S. Narain and T.R. Taylor,
{\it Nucl. Phys.} {\bf B432} (1994) 187.
%
\bibitem{LLEM} A. Lleyda and C. Mu\~noz,
{\it Phys. Lett.} {\bf B317} (1993) 82.
%
\bibitem{CEKLP} D. Choudhury, F. Eberlein, A. K\"oning, J. Louis and 
S. Pokorski, 
{\it Phys. Lett.} {\bf B342} (1995) 180.
%
\bibitem{LNI} J. Louis and Y. Nir, 
{\it Nucl. Phys.} {\bf B447} (1995) 18.
%
\bibitem{BC} P. Brax and M. Chemtob, {\it Phys.Rev.} {\bf D51} (1995) 6550.
%
\bibitem{BS} P. Brax and C.A. Savoy, {\it Nucl. Phys.} {\bf B447}
(1995) 227.
%
\bibitem{LN} For a review, see: A.B. Lahanas and D.V. Nanopoulos,
{\it Phys. Rep.} {\bf 145} (1987) 1, and references therein.
%
\bibitem{CLLM} J.A. Casas, A. Lleyda and C. Mu\~noz,
{\it Nucl. Phys.} {\bf B471} (1996) 3,
{\it Phys. Lett.} {\bf B389} (1996) 305.
%
\bibitem{CH} K. Choi, 
{\it Phys. Rev. Lett.} {\bf 72} (1994) 1592. 
%
\bibitem{BKL} 
B. Acharya, D. Bailin, A. Love, W.A. Sabra and S. Thomas,
{\it Phys. Lett.} {\bf B357} (1995) 387;
D. Bailin, G.V. Kraniotis and A. Love,
%{\it SUSX-TH-}{\bf 97-11},
{\it hep-th/9705244}, {\it hep-th/9707105}.
%
\bibitem{macorra} A. de la Macorra and G.G. Ross, 
{\it Nucl. Phys.} {\bf B404} (1993) 321;
V. Halyo and E. Halyo, {\it Phys. Lett.} {\bf B382} (1996) 89. 
%
\bibitem{BLM} R. Barbieri, J. Louis and M. Moretti,
{\it Phys. Lett.} {\bf B312} (1993) 451 [Erratum: {\bf B316} (1993) 632].
%
\bibitem{LNZ} J.L. Lopez, D.V. Nanopoulos and A. Zichichi,
{\it Phys. Lett.} {\bf B319} (1993) 451.
%
\bibitem{KMV} S. Khalil, A. Masiero and F. Vissani
{\it Phys. Lett.} {\bf  B375} (1996) 154.
%
\bibitem{strings95}
See e.g. L.E. Ib\'a\~nez, {\it hep-th/9505098}, 
Proc. of Strings 95, World Scientific (1995), and references therein.
%
\bibitem{albertus} J.A. Casas, {\it Phys. Lett.} {\bf B384} (1996) 103.
%
\bibitem{BIM2} A. Brignole, L.E. Ib\'{a}\~{n}ez and C. Mu\~noz,
{\it Phys. Lett.} {\bf B387} (1996) 305. 
%
\bibitem{CLLM2} J.A. Casas, A. Lleyda and C. Mu\~noz,
{\it Phys. Lett.} {\bf B380} (1996) 59.
%
\bibitem{KSYY} T. Kobayashi, D. Suematsu, K. Yamada and Y. Yamagishi,
{\it Phys. Lett.} {\bf B348} (1995) 402.
%
\bibitem{BIMS} A. Brignole, L.E. Ib\'{a}\~{n}ez, C. Mu\~noz and C. Scheich,
{\it Z. Phys.} {\bf C74} (1997) 157.
%
\bibitem{gaugino} A. Font, L.E. Iba\~nez, D. L\"ust and F. Quevedo,
{\it Phys. Lett.} {\bf B245} (1990) 401;
S. Ferrara, N. Magnoli, T.R. Taylor and G. Veneziano, 
{\it Phys. Lett.} {\bf B245} (1990) 409;
M. Cvetic, A. Font, L.E. Iba\~nez, D. L\"ust and F. Quevedo,
{\it Nucl. Phys.} {\bf B361} (1991) 194;
B. de Carlos, J.A. Casas and C. Mu\~noz, 
{\it Phys. Lett.} {\bf B299} (1993) 234,
{\it Nucl. Phys.} {\bf B399} (1993) 623;
A. de la Macorra and G.G. Ross,
{\it Phys. Lett.} {\bf B325} (1994) 85.
%
\bibitem{KM} H.B. Kim and C. Mu\~noz,
{\it Z. Phys.} {\bf C75} (1997) 367. 
%
\bibitem{KM2} H.B. Kim and C. Mu\~noz,
{\it Mod. Phys. Lett.} {\bf A12} (1997) 315.
%
\bibitem{IL} L.E. Ib\'{a}\~{n}ez and D. L\"ust,
{\it Nucl. Phys.} {\bf B382} (1992) 305.
%
%\bibitem{coincidencia} Y. Kawamura, T. Kobayashi and J. Kubo,
%{\it DPSU}-{\bf 97-3},
%{\it hep-ph/} {\bf 9703320}.
%
\bibitem{CDG} C.-H. Chen, M. Drees and J.F. Gunion,
{\it Phys. Rev.} {\bf  D55} (1997) 330;
Y. Kawamura, S. Khalil and T. Kobayashi, {\it hep-ph/9703239};
A. Love and P. Stadler, {\it hep-ph/9709234}.
%
\bibitem{KKW} Y. Kawamura, T. Kobayashi and M. Watanabe, 
{\it DPSU-97-5},
{\it hep-ph/9609462}.
%
\bibitem{filq}
A. Font, L.E. Ib\'a\~nez, D. L\"ust and F. Quevedo,
{\it Phys. Lett.} {\bf B249} (1990) 35;
A. Sen, {\it Int.J. Mod.Phys.} {\bf A9} (1994) 3707.
%
\bibitem{mtheory}
See e.g. J.H. Schwarz, {\it hep-th/9607201}, 
P.K. Townsend, {\it hep-th/9612121}, and references therein.
%
\bibitem{W}
E. Witten, {\it Nucl. Phys.} {\bf  B471} (1996) 135.
%
\bibitem{mtpheno}
T. Banks and M. Dine, {\it hep-th/9605136}, {\it hep-th/9608197}, 
{\it hep-th/9609046};
E. Caceres, V.S. Kaplunovsky and I.M. Mandelberg, {\it hep-th/9606036}; 
P. Horava, 
{\it hep-th/9608019}; T. Li, J. Lopez and D.V. Nanopoulos, 
{\it hep-ph/9702237},
{\it hep-ph/9704247}; E. Dudas and C. Grojean, 
{\it hep-th/9704177};
I. Antoniadis and M. Quir\'os, {\it hep-th/9705037},
{\it hep-th/9707208};
K. Choi, {\it hep-th/9706171};
H.P. Nilles, M. Olechowski and M. Yamaguchi, {\it hep-th/9707143};
Z. Lalak and S. Thomas, {\it hep-th/9707223};
V. Kaplunovsky and J. Louis, {\it hep-th/9708049};
E. Dudas, {\it hep-th/9709043}.
%
 


\end{thebibliography}
\end{document}